\documentclass[acmsmall]{acmart}

\AtBeginDocument{%
  }

\setcopyright{rightsretained}
\acmDOI{10.1145/3643773}
\acmYear{2024}
\copyrightyear{2024}
\acmSubmissionID{fse24main-p1198-p}
\acmJournal{PACMSE}
\acmVolume{1}
\acmNumber{FSE}
\acmArticle{47}
\acmMonth{7}
\received{2023-09-29}
\received[accepted]{2024-01-23}

\usepackage[utf8]{inputenc}
\usepackage[T1]{fontenc}
\usepackage{microtype}
\usepackage[para,online,flushleft]{threeparttable}
\usepackage{multirow}

\usepackage{makecell}
\usepackage{subfigure}
\usepackage[skins,breakable]{tcolorbox}

\usepackage{xspace}
\usepackage{soul}
\usepackage{colortbl}
\usepackage{twemojis}

\usepackage[frozencache,cachedir=.]{minted}
\usepackage{color} 
\usepackage{underscore}
\usepackage[normalem]{ulem}
\definecolor{bg}{rgb}{0.95,0.95,0.95} 
\definecolor{grayrow}{rgb}{0.7,0.7,0.7}  

\newminted{markdown}{breaklines, bgcolor=bg,fontsize=\small}
\newminted{diff}{breaklines, bgcolor=bg,fontsize=\small}
\usepackage{graphicx}

\newcommand{\fix}[1]{\textcolor{black}{#1}}

\newcommand{\RqOne}{\textbf{(RQ1)} \ul{\textit{To what extent do developers use Copilot for PRs in the code review process?}}\xspace}

\newcommand{\RqTwo}{\textbf{(RQ2)} \ul{\textit{How are the code reviews affected by the use of Copilot for PRs?}} \xspace} 
\newcommand{\RqTwoDotOne}{\textbf{(RQ2.1)} \ul{\textit{Is there a relationship between the use of Copilot for PRs and review time?}} \xspace} 
\newcommand{\RqTwoDotTwo}{\textbf{(RQ2.2)} \ul{\textit{Is there a relationship between the use of Copilot for PRs and the likelihood of a PR being merged?}} \xspace} 

\newcommand{\RqThree}{\textbf{(RQ3)} \ul{\textit{How do developers adapt the content suggested by Copilot for PRs?}} \xspace} 

\newcommand{\RqThreeDotOne}{\textbf{(RQ3.1)} \ul{\textit{What kind of supplementary information complements the content suggested by Copilot for PRs?}} \xspace}
\newcommand{\RqThreeDotTwo}{\textbf{(RQ3.2)} \ul{\textit{What kind of content suggested by Copilot for PRs undergoes subsequent editing by developers?}} \xspace}
\begin{document}


\title{Generative AI for Pull Request Descriptions: Adoption, Impact, and Developer Interventions}

\author{Tao Xiao}
\orcid{0000-0003-4070-585X}
\affiliation{%
  \institution{Nara Institute of Science and Technology}
  \city{}
  \country{Japan}
}
\email{tao.xiao.ts2@is.naist.jp}

\author{Hideaki Hata}
\orcid{0000-0003-0708-5222}
\affiliation{%
  \institution{Shinshu University}
  \city{}
  \country{Japan}
}
\email{hata@shinshu-u.ac.jp}

\author{Christoph Treude}
\orcid{0000-0002-6919-2149}
\affiliation{%
  \institution{Singapore Management University}
  \city{}
  \country{Singapore}
}
\email{ctreude@smu.edu.sg}

\author{Kenichi Matsumoto}
\orcid{0000-0002-7418-9323}
\affiliation{%
  \institution{Nara Institute of Science and Technology}
  \city{}
  \country{Japan}
}
\email{matumoto@is.naist.jp}

\begin{abstract}
GitHub's Copilot for Pull Requests (PRs) is a promising service aiming to automate various developer tasks related to PRs, such as generating summaries of changes or providing complete walkthroughs with links to the relevant code. As this innovative technology gains traction in the Open Source Software (OSS) community, it is crucial to examine its early adoption and its impact on the development process. Additionally, it offers a unique opportunity to observe how developers respond when they disagree with the generated content. In our study, we employ a mixed-methods approach, blending quantitative analysis with qualitative insights, to examine 18,256 PRs in which parts of the descriptions were crafted by generative AI. Our findings indicate that: (1) Copilot for PRs, though in its infancy, is seeing a marked uptick in adoption. (2) PRs enhanced by Copilot for PRs require less review time and have a higher likelihood of being merged. (3) Developers using Copilot for PRs often complement the automated descriptions with their manual input. These results offer valuable insights into the growing integration of generative AI in software development.
\end{abstract}

\ccsdesc[500]{Social and professional topics~Software maintenance}
\ccsdesc[500]{Software and its engineering~Automatic programming}

\keywords{Pull Requests, Generative AI, Copilot, GitHub}



\maketitle

\section{Introduction}

Generative AI is taking over many areas, and software development is no exception~\cite{ebert2023generative}. As AI becomes more adept at creating content, code, and insights, developers are seeking innovative ways to work alongside these tools~\cite{wang2020human}. It is crucial to understand these interactions to gain insights into the evolution of AI in software development and how developers maintain the quality of software projects despite the AI's involvement.

Pull-based development~\cite{gousios2016work}, due to its widespread use, has seen AI enhancements tailored for its processes. GitHub's Copilot for Pull Requests (PRs) is a prime example~\cite{githubnextGitHubNext}. The tool is specifically designed to elevate the Pull Request experience: aiding developers in crafting optimised PR descriptions and enabling teams to review and merge PRs more efficiently. By tracking developers' work, suggesting tailored descriptions, and providing comprehensive code walkthroughs, it is transforming the review process. Copilot for Pull Requests allows developers to insert specific markers, which it then expands into rich content detailing the changes. For example, while \texttt{copilot:summary} gives a brief overview, \texttt{copilot:walkthrough} provides a detailed list of changes, including direct links to the relevant code. Thousands of PR descriptions have already benefitted from such automation.

With the tool's growing adoption, it is imperative to ask: How does it shape the code review process? Does it expedite reviews? Is there a higher chance of PRs being merged due to its input? Addressing these questions builds upon the existing research on factors influencing code review~\cite{baysal2016investigating}, and the insights could be crucial for developers contemplating the adoption of generative AI tooling.

Copilot for PRs also offers a unique lens into human-AI collaboration in the context of software development~\cite{wu2021ai}. With access to the full edit history of PRs, we can observe how developers adjust AI-generated content. Which details do they add or change? These interactions can reveal not only the dynamics between developers and AI tools but also hint at potential gaps and limitations of AI models in the software development arena.

To investigate these questions, we collected 18,256 PRs powered by Copilot for PRs from 146 GitHub projects, with 1,437 revisions that modified content suggested by Copilot for PRs, as well as 54,188 PRs that are not powered by Copilot for PRs from the same set of GitHub projects.
This study is an exploratory study of early adoption of a new service during its limited release phase, similar to a previous study that examined early adopter usage when GitHub Discussions was in beta and only available to a limited number of users~\cite{10.1007/s10664-021-10058-6}.

Our work answers the following research questions:

\noindent\RqOne

\noindent\RqTwo

\RqTwoDotOne

\RqTwoDotTwo

\noindent\RqThree

\RqThreeDotOne

\RqThreeDotTwo

We have observed (i) a burgeoning adoption of Copilot for PRs during code reviews. Repositories that have been using it for several months have extensively embraced this feature. Notably, \texttt{copilot:summary} stands out as the most popular marker tag, with 13,231 instances; (ii) the use of Copilot for PRs has resulted in a substantial reduction of review time by an average of 19.3 hours, with Pull Requests assisted by Copilot for PRs having a 1.57 times higher likelihood of being merged compared to those not assisted by Copilot for PRs; and (iii) there are 13 categories of supplementary information and seven editorial actions identified that developers employ on Copilot-suggested content. Developers often integrate templates (22.8\%) and add relevant links (22.7\%) to the content suggested by Copilot for PRs, while also frequently partially removing the suggested content (22.9\%).

\section{Background}
In this section, we provide an overview of Copilot for PRs (Section~\ref{sec:bg}) and discuss its context within the broader literature concerning large language models and Pull Request summarization that is pertinent to this study (Section~\ref{sec:rw}).

\subsection{Copilot for PRs}
\label{sec:bg}
With the increasing integration of Large Language Models (LLMs) into a variety of applications, GitHub has announced Copilot for Pull Requests, a tool designed to increase developer efficiency by integrating AI assistance into the code review process. The new service, which is not yet available to the public, extends GitHub's core Pull Requests (PRs) functionality by leveraging AI's ability to write PR descriptions and assist with the review and merge process~\cite{githubnextGitHubNext}.
Among the many features that Copilot for PRs is supposed to enable, there is a feature to generate pull request descriptions, only this feature allows developers to request to be waitlisted for use in a particular repository.\footnote{\url{https://github.blog/2023-03-22-github-copilot-x-the-ai-powered-developer-experience/}}
This feature allows developers to incorporate specific marker tags in their PR descriptions to append AI-generated content, courtesy of the \texttt{GPT-4} model by OpenAI. The marker tags and their corresponding functionalities are as follows:

\begin{itemize}
\item \texttt{copilot:summary}: Generates a concise summary of the changes encompassed in the PR.
\item \texttt{copilot:walkthrough}: Provides a comprehensive list of modifications, each accompanied by links to the pertinent code segments.
\item \texttt{copilot:poem}: Crafts a creative poem encapsulating the essence of the changes.
\item \texttt{copilot:all}: Commands the inclusion of all the content types mentioned above.
\end{itemize}

\begin{figure}[t]
    \centering
    \includegraphics[width=.8\textwidth]{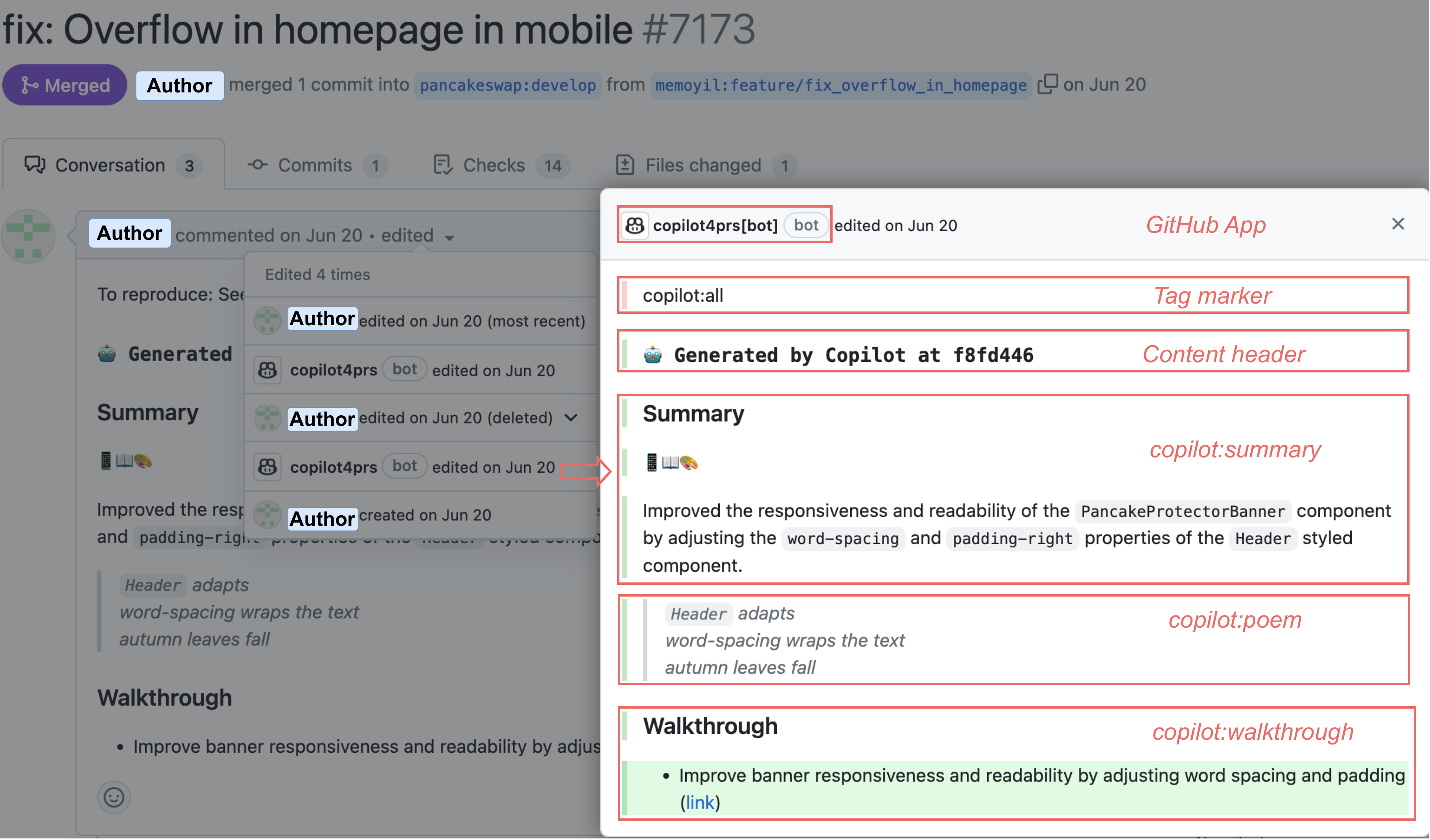}
    \caption{The GitHub Copilot for PRs feature: an example of \texttt{copilot:all}.}
    \label{fig:example}
\end{figure}

We have found that several repositories have already been authorized to use and benefited from the feature of pull request description generation in Copilot for PRs.
Figure~\ref{fig:example} presents an example of the inclusion of the Copilot for PRs marker tag---\texttt{copilot:all}. In this instance, a GitHub App bot, specifically \texttt{copilot4prs}, automatically edits the PR description based on the provided marker tag. The bot generated a summary, a poem, and a walkthrough of the changes in this PR, as depicted in the figure. Developers can review the modified content, make additional adjustments, and even re-include a marker tag for subsequent commits.

\subsection{Related Work}
\label{sec:rw}
\textbf{LLMs for SE.}
Since the introduction of the Transformer architecture in 2017~\cite{vaswani2017attention}, LLMs have gained traction in the field of Software Engineering (SE). \citet{hou2023large} embarked on a systematic review of 229 research articles focusing on the application of LLMs in SE from 2017 to 2023. The review revealed that a significant portion of the studies centered on solving problems in the software development domain. In this domain, LLMs, particularly \texttt{GPT-2/GPT-3/GPT-3.5}~\cite{dong2023self, li2023enabling,liu2023improving,liu2023your,nascimento2023comparing,wang2023evaluating,yeticstiren2023evaluating}, \texttt{GPT-4}~\cite{bareiss2022code,gilbert2023semantic,jiang2023selfevolve,liu2023your}, and the \texttt{BERT} series~\cite{zeng2022extensive,lai2023ds}, have shown a notable efficacy in tasks like code generation, code completion, and code summarization.

Code completion is an integral feature in Integrated Development Environments (IDEs) and code editors. Tools like \texttt{Codex}~\cite{chen2021evaluating, doderlein2022piloting, li2023cctest, pearce2023examining}, \texttt{BERT} series~\cite{khan2022automatic}, \texttt{GitHub Copilot}~\cite{doderlein2022piloting, li2023cctest, pudari2023copilot}, \texttt{CodeParrot}~\cite{xu2022systematic,li2023cctest}, and \texttt{GPT} series~\cite{xu2022systematic,ochs2023evaluating} offer intelligent and accurate code suggestions, greatly enhancing the coding process. In contrast, code summarization technologies like \texttt{Codex}~\cite{ahmed2023improving, arakelyan2023exploring,gao2023constructing}, \texttt{CodeBERT}~\cite{chen2022transferability,gu2022assemble,gao2023constructing}, and \texttt{T5}~\cite{mastropaolo2021studying, mastropaolo2022using} focus on generating natural language descriptions from source code, promoting enhanced code maintenance, search, and classification.

Another application of LLMs in the software maintenance domain---accounting for nearly a quarter of the studies reviewed by \citet{hou2023large}---covers program repair, code review, and debugging. In the context of program repair, \texttt{Codex}~\cite{wu2023effective,xia2023automated} and \texttt{ChatGPT}~\cite{xia2023conversational} have exhibited strong performance. For code review tasks, LLMs like \texttt{BERT}~\cite{sghaier2023multi} and \texttt{ChatGPT}~\cite{sridhara2023chatgpt} are instrumental in accurately detecting issues and suggesting code optimizations. Debugging, the process of identifying, locating, and resolving bugs, has also been enhanced by LLMs, with notable contributions from \texttt{AutoSD}~\cite{kang2023explainable} and \texttt{SELF-DEBUGGING}~\cite{chen2023teaching}.

\textbf{PR summarization.}
PR summarization is closely related to Copilot for PRs. \citet{liu2019automatic} introduced \texttt{Attn}~\cite{bahdanau2014neural}, an attentional encoder-decoder model dubbed \texttt{PRSummarizer}, which leverages features from commit messages and added code comments to automatically generate PR descriptions. This model, assessed using the \texttt{ROUGE} metric~\cite{lin2004rouge} and human evaluation, surpassed the performance of two baselines: \texttt{LeadCM} and \texttt{LexRank}~\cite{erkan2004lexrank}. Enhancing this work, \citet{fang2022prhan} proposed \texttt{PRHAN}, a novel hybrid attention network that effectively addresses the low efficiency and out-of-vocabulary (OOV) challenges identified in the model by \citet{liu2019automatic}, delivering superior results.

Additionally, \citet{irsan2022autoprtitle} delved into the realm of PR title generation, evaluating several summarization tools, including general-purpose models like \texttt{BERTSumExt}~\cite{liu2019text}, \texttt{BART}~\cite{lewis2019bart}, and \texttt{T5}~\cite{raffel2020exploring}, as well as domain-specific models like \texttt{PRSummarizer}~\cite{liu2019automatic} and \texttt{iTAPE}~\cite{chen2020stay}. Their assessments, based on the \texttt{ROUGE} metric and human evaluation, identified \texttt{BART} as the most effective tool for PR title generation. Their fine-tuned model, \texttt{AUTOPRTITLE}, which incorporates additional features, outperformed the others. 

While previous research primarily concentrates on the development and evaluation of LLMs in code generation, completion, summarization, and maintenance tasks, our study is uniquely positioned in exploring the real-world applicability and impact of these models in a practical software development environment. We investigate Copilot for PRs, analyzing its impact on review time and merge decisions, and the edits made to the generated content. This holistic approach provides a comprehensive insight into the benefits and potential challenges associated with the integration of LLMs in everyday software engineering practices.

\section{Data Collection}
In this section, we outline our procedure for collecting data on PRs generated by Copilot for PRs (Section~\ref{sec:treat}), PRs not generated by Copilot for PRs (Section~\ref{sec:control}), and the revisions of PRs generated by Copilot for PRs (Section~\ref{sec:edit}).

\subsection{PRs Generated by Copilot for PRs}
\label{sec:treat}

To investigate Copilot's effectiveness in summarizing PRs, we began by compiling a list of PRs generated by Copilot for PRs. Utilizing GitHub GraphQL search,\footnote{\url{https://docs.github.com/en/graphql/reference/queries\#search}} we identified PRs containing the phrase ``Generated by Copilot'' in their descriptions, created on or before 31st August 2023. \fix{To manage the limitation of GitHub's maximum of 1000 responses per query, we implemented a strategy of dividing our search criteria (creation time in this case). Specifically, when a query for PRs within a certain time period yielded more than 1000 results, we halved the time period and repeated the search. This process was continued until the number of PRs fell within the 1000 result limit, ensuring we could efficiently gather all relevant data without exceeding GitHub's response constraints.}
We excluded false positives where developers included this phrase,\footnote{For example, \url{https://github.com/Kudoser/SteamTogether/pull/17}} retaining only those PRs edited by \texttt{copilot4prs}, a GitHub bot for Copilot for PRs. This resulted in a total of 18,858 PRs across 150 GitHub repositories. 

\textbf{Identifying Obsolete Uses of Copilot for PRs.} 
In our examination of PRs, we noted instances where developers tested Copilot for PRs on old PRs.\footnote{\url{https://github.com/argoproj/argo-cd/pull/1129}} To maintain dataset integrity, we excluded PRs that were (i) closed before being edited by `copilot4prs', and (ii) created before Copilot for PRs was introduced on GitHub (\texttt{2023-03-22 17:44:28+00:00}). This left us with 18,322 non-obsolete PRs.

\textbf{Excluding PRs Submitted by Bots.}
We drew from the extensive study by \citet{golzadeh2022accuracy} on bot detection techniques. Implementing two highly accurate methods, the ``bot'' suffix and the list of bots---based on 527 confirmed bots (e.g., \texttt{googlebot}) identified by \citet{golzadeh2021ground}---we filtered out bot-submitted PRs. \fix{Moreover, including comments generated by bots could skew the representation of actual reviewer participation in code review discussions. To maintain the integrity and accuracy of our analysis, we focus exclusively on comments from human participants. Therefore w}e also removed comments generated by bots to enhance our dataset's quality. Consequently, we were left with 18,256 valid PRs from 146 GitHub repositories. \fix{The age of these 146 early adopters varies from 55 days (a forked repository for the purpose of exploring Copilot for PRs) to 4,762 days (\texttt{scikit-learn/scikit-learn}---a popular Python module for machine learning), with the average age of 1,360 days around four years.  }

\subsection{PRs Not generated by Copilot for PRs}
\label{sec:control}
To enable a comparative analysis, we also collated a dataset of PRs that were not generated by Copilot for PRs, sourced from the same 146 GitHub repositories. This approach ensures a balanced and fair comparison. Since review times for PRs can vary across different time periods, influenced by factors such as community growth, we selected PRs created from the introduction of the Copilot for PRs feature on GitHub to 31st August 2023 in each repository. This approach aims to control for temporal variations in review times. Like the Copilot-generated PRs, we applied a filtering criterion to exclude PRs submitted by bots to maintain consistency in data quality. This process yielded 54,188 PRs from 139 repositories for our comparative analysis (seven repositories were exclusively comprised of Copilot-generated PRs post-filtering during this specific timeframe). Table~\ref{tab:state} presents the final count of PRs analyzed in this study. Notably, our initial observations indicate that the acceptance rate for Copilot-generated PRs (84\%) is superior to that for non-Copilot-generated PRs (71\%).

\begin{table}[t]
    \centering
    \caption{The distribution of the state of studied PRs.}
    \label{tab:state}
    \begin{tabular}{lr@{}rr@{}r}
    \toprule
& \multicolumn{2}{c}{\textbf{Copilot-generated PRs}} & \multicolumn{2}{c}{\textbf{non-Copilot-generated PRs}} \\ 
     \midrule
      \# merged  &   15,270 & (84\%) &  38,639 & (71\%) \\
      \# closed   &  1,907 & (10\%) &  12,056 & (22\%) \\
      \# opened   &  1,079 & (6\%) &  3,493 & (6\%) \\
      \midrule
     \textbf{sum}  & 18,256  & (100\%) &  54,188 & (100\%) \\
    \bottomrule
    \end{tabular}
\end{table}

\subsection{Revisions of PR Descriptions Generated by Copilot for PRs}
\label{sec:edit}
We investigate how developers engage with the content suggested by Copilot for PRs (\textbf{RQ3}) by obtaining revisions of PR descriptions from 17,177 merged/closed PRs generated by Copilot for PRs.

\textbf{Collection of Edits on PR Descriptions.}
We gather data on the edits made to each PR description, including information about who made the edit, the content changes, and the time of the edit. As noted in Section~\ref{sec:bg}, \texttt{copilot4prs} serves as the editor, replacing marker tags with content suggested by Copilot for PRs. We collected 46,700 revisions from 17,177 merged/closed PRs generated by Copilot for PRs.

\textbf{Exclusion of PRs Without Post-Copilot Edits.}
Our focus is on understanding how developers adopt and adapt suggestions from Copilot for PRs during the code review process. Therefore, we exclude PRs that have not been edited following the initial placement of marker tags. This criteria yielded 18,486 revisions from 3,935 PRs.

\textbf{Filtering PRs that Reapply Marker Tags.}
In our preliminary analysis of PRs generated by Copilot for PRs, we noticed instances where developers reapplied the same marker tag to examine the newly generated content after adding new commits. For instance, in this PR,\footnote{\url{https://github.com/ultralytics/ultralytics/pull/1956}} the author reapplied the \texttt{copilot:all} tag after adding commit \#143ee5c. After applying this filter, we were left with 4,391 revisions from 730 PRs.

\textbf{Identification of PRs with Post-Copilot Edits.}
We employ the \texttt{git diff} (Myers algorithm) on each content of revisions in PRs to trace modifications made to the content initially generated by \texttt{copilot4prs}. This process allowed us to identify 1,437 revisions spanning 311 PRs where developers made additional edits.

\textbf{Identifying the PR Template.}
To segregate information inherent in the PR template, which is present prior to PR creation, we use the GitHub GraphQL API\footnote{\url{https://docs.github.com/en/graphql/reference/objects\#pullrequesttemplate}} to retrieve the PR template. We then pinpoint the version of the PR template closest to the PR creation time using \texttt{git log}. In cases where multiple PR templates could potentially auto-generate PR content, we calculate pairwise cosine similarities between each template and the initial PR description to identify the most similar template. These templates are subsequently used for answering \textbf{RQ3.1}.

\section{Methods}
We present our mixed-methods procedure, comprising a quantitative analysis (\textbf{RQ1}), two quasi-experiments for causal inference (\textbf{RQ2}), and a qualitative analysis (\textbf{RQ3}).

\subsection{Quantitative Analysis}
\RqOne\\
To tackle \textbf{RQ1}, our quantitative examination \fix{of 18,256 valid PRs from 146 GitHub repositories} emphasizes three elements related to the application of Copilot for PRs: (1) its adoption trajectory, (2) the extent of its employment in the code review mechanism, and (3) the dispersion patterns of Copilot for PRs marker tags.

\textbf{Adoption Trend of Copilot for PRs.} 
In order to understand the evolution in the acceptance of Copilot for PRs, we study the number of PRs that incorporated this service. Our exploration of its acceptance is limited to the initial phase of its existance, spanning from March 2023 to August 2023.

\textbf{Proportions of Copilot for PRs.} 
To assess the degree to which developers rely on this feature during code reviews, we compute the ratios of Copilot for PRs utilization, represented as $\frac{\# Copilot PRs}{\# PRs}$, within individual GitHub repositories using Copilot for PRs. Subsequently, a bubble plot is constructed to illustrate the temporal evolution of Copilot for PRs engagement across GitHub.

\textbf{Distribution of Marker Tags from Copilot for PRs.}
Copilot for PRs introduces four unique marker tags\footnote{\url{https://web.archive.org/web/20231023053319/https://github.com/apps/copilot4prs}} enabling customization of the generated content. Developers can incorporate these tags in their PR descriptions to elicit diverse responses from Copilot for PRs. 
To quantify the popularity of these marker tags, we use the regular expression for the extraction of their instances from the descriptions of PRs generated by Copilot for PRs.
\begin{verbatim}
\bcopilot:(all|summary|walkthrough|poem)\b
\end{verbatim} 

\subsection{Casual Inference}
\RqTwo\\
\indent\RqTwoDotOne\\
\indent\RqTwoDotTwo\\
To answer these RQs, we build two quasi-experiments to estimate the causal impact of
Copilot for PRs on reducing code review time in Pull Requests (\textbf{RQ2.1}) and increasing the likelihood of a Pull Request being merged (\textbf{RQ2.2}).

\begin{table*}[t]
\centering
\caption{The studied variables in \textbf{RQ2}.}
\label{tab:variables}
\resizebox{\textwidth}{!}{
\begin{tabular}{p{8cm}lrr}
\toprule
\rowcolor{grayrow}
\textbf{PR variables}       & \textbf{Description}          &  \makecell{\textbf{Median of}\\\textbf{Treatment}}          &  \makecell{\textbf{Median of}\\\textbf{Control}}                   \\ 
\# Added lines~\cite{wang2021understanding}                      & The number of added LOC by a PR.  &  \textbf{28} &   25                            \\
\# Deleted lines~\cite{wang2021understanding}                   & The number of deleted LOC by a PR.  &   \textbf{8} &      7                       \\
PR size~\cite{thongtanunam2016revisiting,thongtanunam2017review,kononenko2018studying,mcintosh2016empirical,wang2021understanding}               & The total number of added and deleted LOC by a PR. & \textbf{44} &   41                \\
Purpose~\cite{mcintosh2016empirical,thongtanunam2017review,wang2021understanding}              & The purpose of a PR, i.e., bug, document, and feature.    & - & -              \\

\# Files~\cite{tsay2014influence,thongtanunam2017review,kononenko2018studying,wang2021understanding}                & The number of files changed by a PR.    & \textbf{3} & 2           \\
\# Commits~\cite{tsay2014influence,kononenko2018studying,wang2021understanding}                 & The number of commits involved in a PR.    & \textbf{2} & \textbf{2}           \\
Description length~\cite{thongtanunam2017review,mcintosh2016empirical,wang2021understanding}            & The length of a PR description.   &  \textbf{1,825} &   343         \\
PR author experience~\cite{thongtanunam2017review,kononenko2018studying,wang2021understanding}  & The number of prior PRs that were submitted by the PR author.  & \textbf{155} & 151  \\
Is member~\cite{tsay2014influence, zhang2022pull}   & Whether or not the PR author is a member or outside collaborator (binary).  & - & -  \\
\# Comments~\cite{kononenko2018studying,wang2021understanding}              & The number of comments left on a PR.   & 0 & \textbf{1}                   \\
\# Author comments~\cite{kononenko2018studying,wang2021understanding}       & The number of comments left by the PR author.  & \textbf{0} & \textbf{0}       \\
\# Reviewer comments~\cite{kononenko2018studying,wang2021understanding}     & The number of comments left by the reviewers who participate in the discussion. & \textbf{0} & \textbf{0}            \\
\# Reviewers~\cite{thongtanunam2016revisiting,wang2021understanding} & The number of developers who participate in the discussion.   & \textbf{0} & \textbf{0}           \\
Repo age~\cite{tsay2014influence,zhang2022pull}               & Time interval between the repository creation time and PR creation time in days.     & 1,060 & \textbf{1,250}                \\
\midrule
\rowcolor{grayrow}
\textbf{Project variables}       &           &            &                      \\
Language~\cite{zhang2022pull}                & The repository language that a PR belongs to, represented by the top 10 or others.     & - & -                \\
\# Forks~\cite{zhang2022pull}                & The number of forks that a repository has.    & 264 & \textbf{286}                \\
\# Stargazers~\cite{tsay2014influence,zhang2022pull}              & The number of stargazers that a repository has.    & 1,359 & \textbf{1,654}                \\
\midrule
\rowcolor{grayrow}
\textbf{Treatment variables} & & & \\
With Copilot for PRs    & Whether or not a PR is generated by Copilot for PRs (binary). & - & - \\
\midrule
\rowcolor{grayrow}
\textbf{Outcome variables} & & & \\
Review time \textbf{(RQ2.1)}    &  Time interval
between the PR creation time and closed time in hours. & 12.17 & \textbf{16.09}\\
Is merged \textbf{(RQ2.2)}    & Whether or not a PR is merged (binary).  & - & -\\
\bottomrule
\end{tabular}}
\end{table*}

We define the review time as the time interval
between the PR creation date and closed date in hours. In our analysis, we employ a statistical adjustment technique known as the Propensity Score Weighting (PSW) method to account for potential confounding factors in our observational data. This technique serves to calculate the inverse of the propensity score as a weight applied to each unit in the treatment group and the inverse of one minus the propensity score as a weight for each unit in the control group~\cite{rubin2001using}. The propensity score itself is defined as the conditional probability of receiving treatment given a set of observed covariates~\cite{rosenbaum1983central}. Through this weighting scheme, the PSW method aims to construct a balanced pseudo-population in which the distribution of observed covariates is equivalent across both treatment and control groups~\cite{rubin2001using,rosenbaum1983central}. In this analysis, the 17,177 (15,270 + 1,907) merged/closed PRs generated by
Copilot for PRs as the treatment group, and the 50,695 (38,639 + 12,056) merged/closed PRs not generated by
Copilot for PRs as the control group.

\textbf{Explanatory Variables.} Table~\ref{tab:variables} presents the 18 metrics that are used as explanatory variables in the logistic
regression to estimate the propensity score for PSW. The treatment variable is whether Copilot for PRs is used to generate the PR description. The other 17 variables have been shown to have an impact on the review time, outcome, quality, and review participation of the modern code review process in previous studies~\cite{tsay2014influence,thongtanunam2016revisiting,thongtanunam2017review,kononenko2018studying,mcintosh2016empirical,wang2021understanding,zhang2022pull}. Similar to the prior work~\cite{mockus2000identifying, mcintosh2016empirical,thongtanunam2017review,wang2021understanding}, we classify the purpose of a PR for which the description contains ``doc'', ``copyright'' or ``license'' as documentation, and if a PR description contains ``fix'', ``bug'', or ``defect'', it is classified as bug fixing. The remaining PRs are classified as feature introduction. We adopt the fork count and stargazer count as project variables since they are indications of the attention a project receives. Moreover, different primary programming languages of projects may also infer the review time.
Additionally, we identified whether or not the PR author is a member or outside collaborator according to the GitHub GraphQL.\footnote{\url{https://docs.github.com/en/graphql/reference/enums\#commentauthorassociation}} We count `MEMBER' and `OWNER' as members of a GitHub repository.

\begin{figure}
  \centering
  
\includegraphics[width=.8\linewidth]{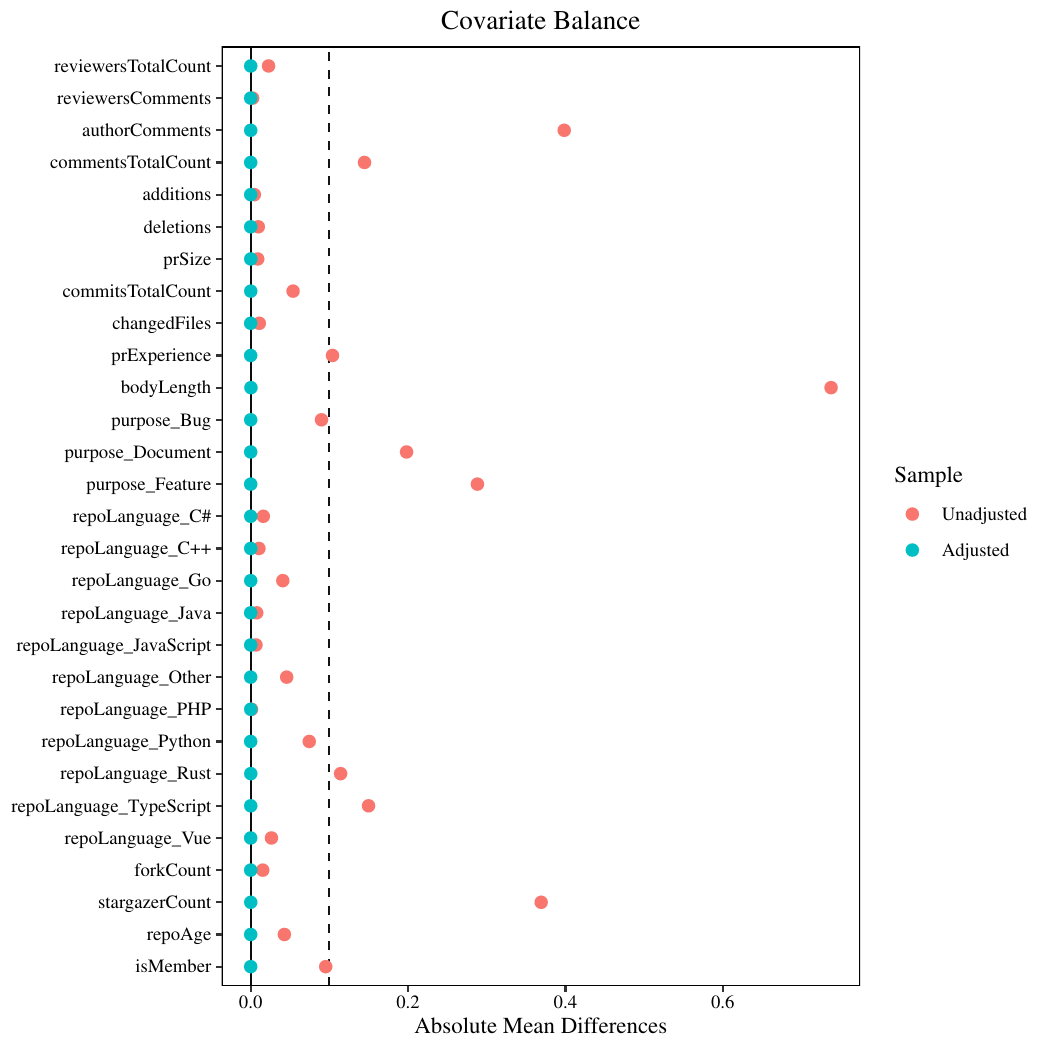}
  \caption{Covariate balance before (unadjusted) and after (adjusted) propensity score weighting.}
  \label{fig:balance}
\end{figure}

\textbf{Entropy Balancing.} 
We employed the Entropy Balancing method~\cite{hainmueller2012entropy}, which ensures optimal balance for specified statistical moments of the covariates while minimizing the entropy of the weights. \fix{The reasoning behind this choice is that Entropy Balancing offers a more flexible and efficient way to reweight units while retaining valuable information in the preprocessed data, preventing the need for continual balance checking and iterative searching over propensity score models, which may fail to balance covariate distributions in finite samples~\cite{hainmueller2012entropy}. }
Figure~\ref{fig:balance} illustrates the reduction in absolute mean differences as a result of applying this method, transitioning from unadjusted to adjusted states (where `unadjusted' encompasses all PRs prior to balancing, and `adjusted' refers to those post-balancing). None
of the absolute mean differences of adjusted exceeds 0.10, which
means that we obtained weighted variables for the treatment and
control groups with a balanced distribution of covariates.

\textbf{Treatment and Outcome Variable.}
To estimate the impact of Copilot for PRs usage in code reviews, a linear regression is performed using the above variables and the variable \texttt{treatment}, which takes a value of 0 or 1 that indicates the presence or absence of Copilot for PRs.
On the one hand, the outcome variable for \textbf{RQ2.1} is the review time, defined as the time interval between the PR creation date and the date it was closed, measured in hours. On the other hand, the outcome variable for \textbf{RQ2.2} is whether or not a PR is merged, where we estimate the causal impact of the use of Copilot for PRs on the likelihood of a PR being merged.

\subsection{Qualitative Analysis}
\RqThree\\
For the qualitative analysis examining how developers complement and modify content suggested by Copilot for PRs, we qualitatively analysed 1,437 revisions from 311 PRs where developers made additional edits \fix{as illustrated in Section \ref{sec:edit}}. 
To increase confidence in our qualitative processes, three of the
authors initially collaboratively reviewed a set of 30 samples. This preliminary examination focused on identifying recurring themes relevant to our research questions. Following this, one author formalized these discussions into a well-defined coding schema. Subsequently, three authors employed this coding schema to analyze another 30 samples. One author then finalized the annotations based on the encouraging kappa agreements. We allowed multiple codes per PR. In all cases, the kappa agreement proved satisfactory on the first pass, obviating the need for further revisions.

\RqThreeDotOne
To answer \textbf{RQ3.1}, we analyzed how developers complement the content suggested by Copilot for PRs. Three raters
independently coded 30 samples, achieving a kappa
of 0.64 or `substantial' agreement~\cite{viera2005understanding}. The lower agreement can be
explained by 24 combinations of multiple codes being discovered
when coding complementary information to the content suggested by Copilot for PRs. Three raters achieved perfect
agreement in 15/30 cases (50\%), partial agreement in another
14/30 cases (47\%), and completely disagreed only in 1/30 case (3\%).
Our coding schema and the frequencies of different codes are shown in Table~\ref{tab:code1}. 

\RqThreeDotTwo
To answer \textbf{RQ3.2}, we analyzed how developers modified the content suggested by Copilot for PRs. Three raters
independently coded 30 samples, achieving a kappa
of 0.62 or `substantial' agreement~\cite{viera2005understanding}. The lower agreement can be
explained by 26 combinations of multiple codes being discovered
when coding modified information of the content suggested by Copilot for PRs. Three raters achieved perfect
agreement in 14/30 cases (47\%), partial agreement in another
15/30 cases (50\%), and completely disagreed only in 1/30 case (3\%).
Our coding schema and the frequency of different codes are shown in Table~\ref{tab:code2}. 

\section{Results}
In this section, we present the results of our research questions 1--3. 

\subsection{RQ1: To what extent do developers use Copilot for PRs in the code review process?}

\begin{figure}[t]
    \centering

    \includegraphics[width=.8\textwidth]{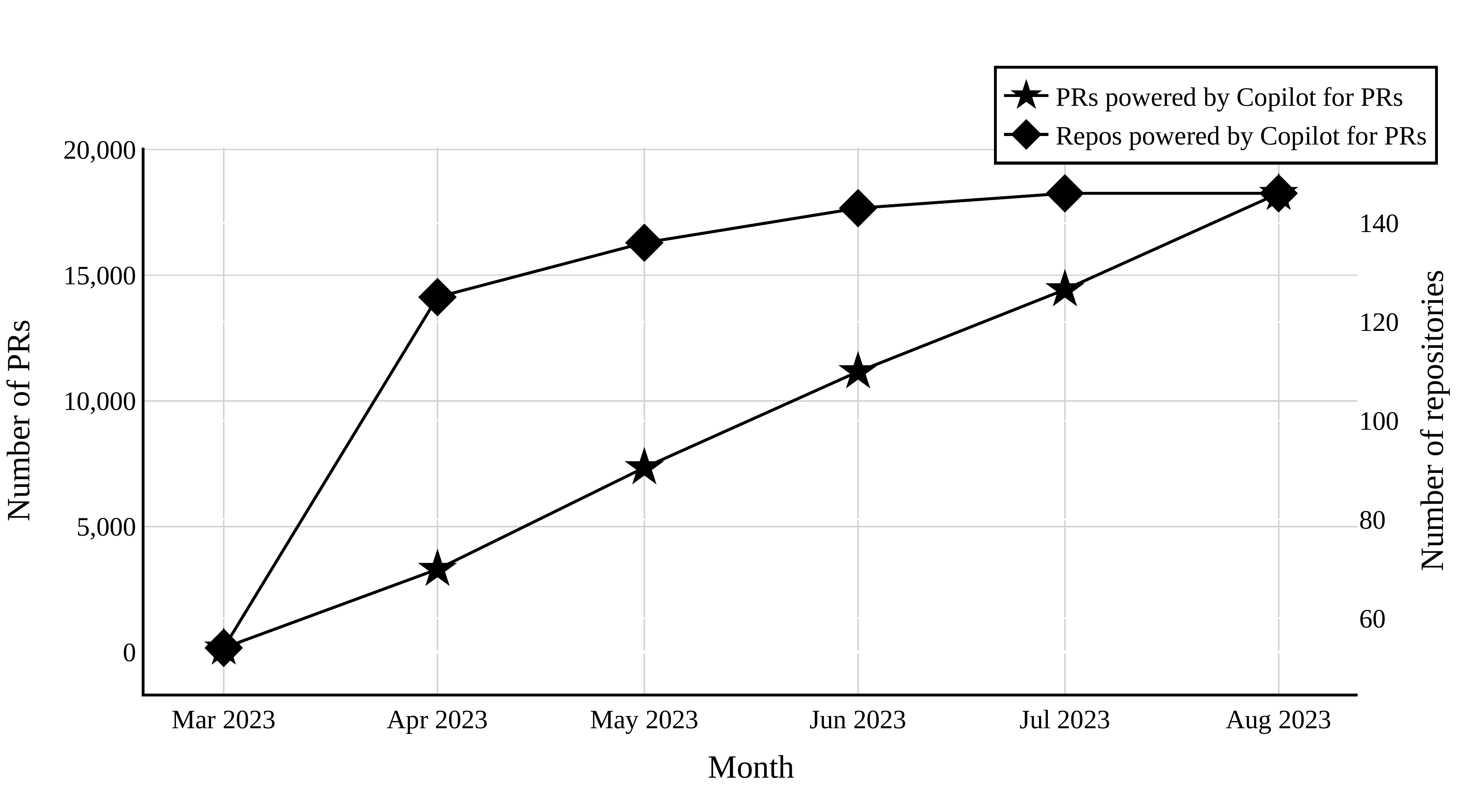}
    \caption{Cumulative time-series of PRs using Copilot for PRs vs. Non-Copilot for PRs.}
    \label{fig:RQ1.1}
\end{figure}

\begin{figure}[t]
    \centering
    \includegraphics[width=.8\textwidth]{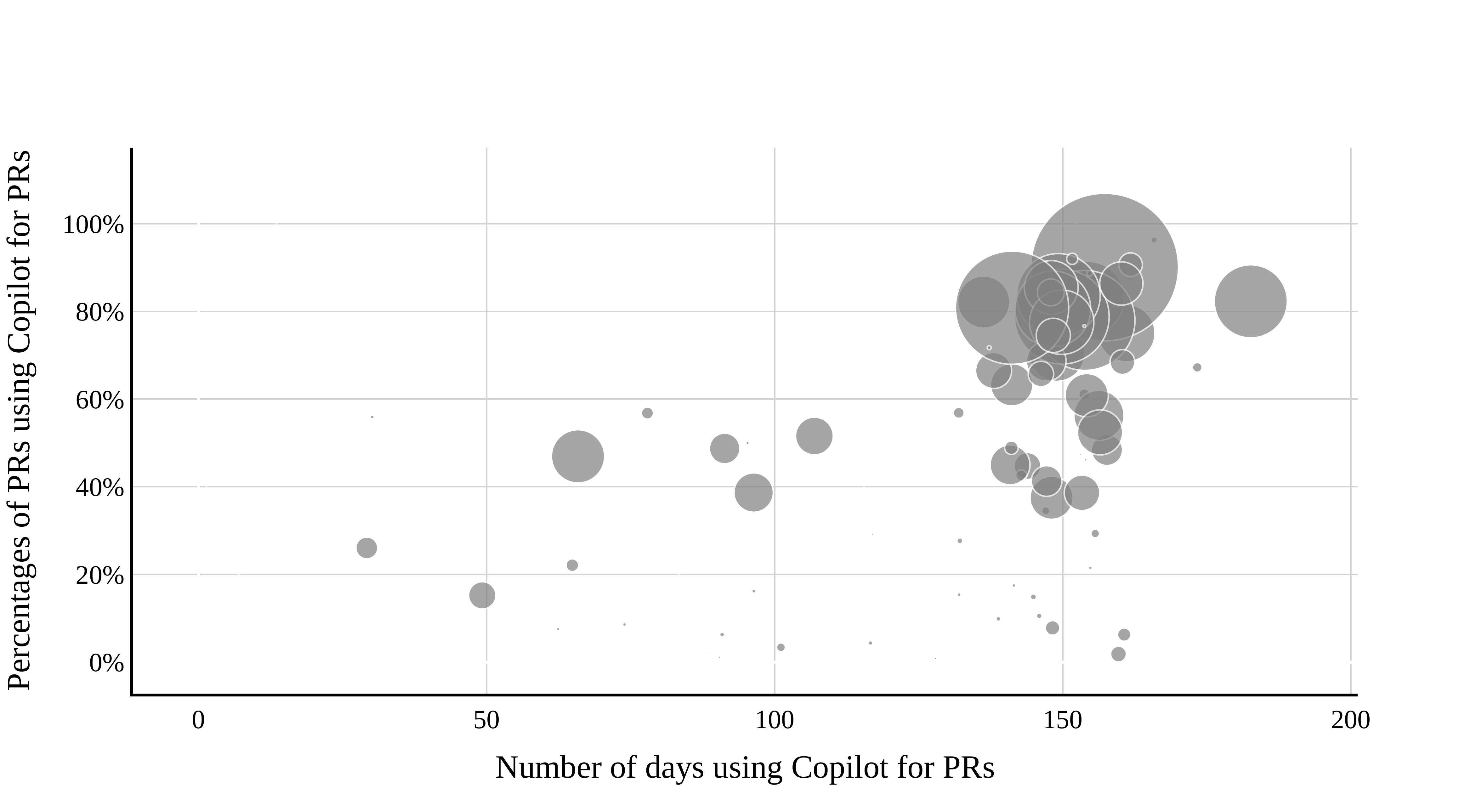}
    \caption{Proportions of PRs using Copilot for PRs per repository.}
    \label{fig:RQ1.2}
\end{figure}

Figures~\ref{fig:RQ1.1}--\ref{fig:RQ1.2} and Table~\ref{tab:RQ1} show the results of our analysis. We now discuss our results below.

\textbf{Adoption Trend of Copilot for PRs.} Figure~\ref{fig:RQ1.1} presents the cumulative time-series of PRs using Copilot for PRs. We began collecting PRs created after the initial instance of a PR in GitHub featuring Copilot for PRs, which occurred on \texttt{2023-03-22 17:44:28+00:00}. 
The number of repositories using Copilot for PRs has barely increased since the first month. It seems that only a limited number of repositories are accepted as early adopters of this feature. On the other hand, the number of pull requests containing automatically generated descriptions by Copilot for PRs has been steadily increasing. This suggests that early adopters are actively using this new feature.

\textbf{Proportions of Copilot for PRs.} Figure~\ref{fig:RQ1.2} depicts the bubble plot of the proportions of PRs using Copilot for PRs across repositories. In the bubble plot, the y-axis represents the rates of PRs utilizing Copilot for PRs, while the size of each bubble corresponds to the number of PRs in that repository using Copilot for PRs. The x-axis showcases the duration for which Copilot for PRs has been employed in the repository. From the bubble plot, we can see that repositories with long experience periods (x-axis) with Copilot for PRs have many PRs with Copilot for PRs (bubble size is bigger) and its ratio (y-axis) is also high. Notably, 50 repositories have a heavy reliance on Copilot for PRs, with over 50\% of their PRs being powered by it. In more extreme cases, seven repositories have wholly incorporated Copilot for PRs into their PR descriptions or their PR description templates, a topic further explored in \textbf{RQ3}. Conversely, 96 repositories only sporadically employ this feature, i.e., rates less than or equal to 50\% of PRs.

\begin{table}[t]
    \centering
    \caption{Distribution of marker tags in PRs generated by Copilot for PRs.}
    \begin{tabular}{llr}
    \toprule
    \multirow{4}{*}[-1.6em]{\rotatebox[origin=c]{90}{Marker Tag}} &  \textbf{Category} & \textbf{\#} \\
    \midrule
    & copilot:summary & 13,231 \\
    & copilot:walkthrough & 8,990 \\
    & copilot:summary &  4,952\\
    & copilot:poem & 4,206 \\
    \midrule
    \multirow{10 }{*}[-3.5em]{\rotatebox[origin=c]{90}{Combinations of Marker Tags}}& copilot:summary and copilot:walkthrough & 5,598 \\
    &copilot:all &  4,512\\
    &copilot:summary & 3,725 \\
    &copilot:summary, copilot:poem, and copilot:walkthrough & 2,772 \\
    &copilot:summary and copilot:poem & 716 \\
    &copilot:all, copilot:summary, copilot:poem, and copilot:walkthrough & 412 \\
    &copilot:poem & 275 \\
    &copilot:walkthrough &  195\\
    &copilot:all and copilot:poem & 19 \\
    &none & 14 \\
    &copilot:poem and copilot:walkthrough & 9 \\
    &copilot:all, copilot:summary, and copilot:poem   & 3 \\
    &copilot:all, copilot:summary, and copilot:walkthrough & 3 \\
    &copilot:all and copilot:summary & 2 \\
    &copilot:all and copilot:walkthrough  & 1 \\
    \bottomrule
    \end{tabular}
    \label{tab:RQ1}
\end{table}

\textbf{Distribution of Marker Tags in Copilot for PRs.}
We obtained a total of 31,379 instances of marker tags from 18,256 PRs generated by Copilot for PRs. The top portion of Table~\ref{tab:RQ1} shows the distribution of marker tags in PRs generated by Copilot for PRs, while the bottom portion of Table~\ref{tab:RQ1} presents the combinations of marker tags in these PRs. We find that the \texttt{copilot:summary} tag is the most frequently occurring, accounting for 13,231 instances, while \texttt{copilot:poem} emerges as the least frequent.
The four most common combinations of marker tags are associated with Summary and Walkthrough content (5,598 PRs), only the All content (4,512 PRs), only the Summary content (3,725 PRs), and a combination of Summary, Poem, and Walkthrough content (2,772 PRs). We discovered 14 PRs with no marker tags in their descriptions, as these commented-out marker tags (we identified marker tags by the commented-out contents in PR descriptions) were removed by developers.

\begin{tcolorbox}
\textbf{RQ1 Summary:} 
We observe the growing adoption of Copilot for PRs in the code review process. While a substantial portion of repositories have adopted this feature extensively, over half utilize it only minimally. The most commonly employed Copilot for PRs marker tag is \texttt{copilot:summary}, with 13,231 instances. The most popular combination is associated with  \texttt{copilot:summary} and \texttt{copilot:walkthrough}, accounting for 5,598 PRs.

\end{tcolorbox}

\subsection{RQ2: How are the code reviews affected by the use of Copilot for PRs?}

\begin{table}[t]
    \centering
    \caption{Summaries of causal inference estimating the effect of Copilot for PRs.}
    \label{tab:rq2}
    \begin{threeparttable}
    \begin{tabular}{llll}
    \toprule
    & \textbf{estimate} & \textbf{std. error} & \textbf{p} \\
    \midrule

\textbf{treatment} & \textbf{-19.3} & 2.27 & \textbf{1.64e-17} \\
reviewersTotalCount & 15.0 & 1.59 & 4.17e-21 \\
reviewersComments & 0.996 & 0.836 & 2.34e-1 \\
authorComments & 38.8 & 1.26 & 3.21e-206 \\
commentsTotalCount & NA & NA & NA \\
additions & 0.0000266 & 0.0000198 & 1.79e-1 \\
deletions & -0.000000194 & 0.0000413 & 9.96e-1 \\
prSize & NA & NA & NA \\
commitsTotalCount & 0.753 & 0.0431 & 4.03e-68 \\
changedFiles & -0.0167 & 0.00879 & 5.76e-2 \\
prExperience & -0.0503 & 0.00288 & 4.40e-68 \\
bodyLength & 0.00261 & 0.0000802 & 8.79e-231 \\
purposeDocument & 2.12 & 2.59 & 4.12e-1 \\
purposeFeature & -0.623 & 2.89 & 8.30e-1 \\
repoLanguageC++ & -49.5 & 8.71 & 1.32e-8 \\
repoLanguageGo & -66.2 & 6.16 & 6.36e-27 \\
repoLanguageJava & -46.2 & 7.99 & 7.30e-9 \\
repoLanguageJavaScript & -64.5 & 6.85 & 4.79e-21 \\
repoLanguageOther & -52.3 & 6.23 & 5.24e-17 \\
repoLanguagePHP & -9.75 & 9.02 & 2.80e-1 \\
repoLanguagePython & 21.0 & 6.64 & 1.58e-3 \\
repoLanguageRust & -34.3 & 7.35 & 3.21e-6 \\
repoLanguageTypeScript & -53.5 & 5.56 & 7.66e-22 \\
repoLanguageVue & -44.9 & 7.81 & 9.01e-9 \\
forkCount & -0.000364 & 0.000276 & 1.87e-1 \\
stargazerCount & -0.0000786 & 0.000118 & 5.05e-1 \\
repoAge & 0.0190 & 0.00118 & 1.16e-58 \\
isMember & -28.5 & 2.27 & 3.85e-36 \\

    \bottomrule
    \end{tabular}
    \begin{tablenotes}
\small
\textit{NA} indicates the Perfect Collinearity, where this variable can be predicted by other variables.
\end{tablenotes}
    \end{threeparttable}
\end{table}

\RqTwoDotOne
Table~\ref{tab:variables} presents the studied variables and their median values in the treatment group and control group. We observe that the median value of outcome variables (i.e., review time) of the treatment group (i.e., 12.17 hours) is less than the control group (i.e., 16.09 hours).
Table~\ref{tab:rq2} summarizes the regression result. As seen in the coefficient estimate of \texttt{treatment}, there is a statistically significant positive effect of Copilot for PRs on reducing code review time in RPs.
As the average of the expected causal effect of treatment on individuals in the treatment group, called Average Treatment Effects on the Treated (ATT), we find that Copilot for PRs has an impact of decreasing the review time by 19.3 hours. 

\RqTwoDotTwo 
To compute the marginal log odd ratio (OR) for a PR being merged, we employ \texttt{avg\_comparisons()} with the compassion of \texttt{lnoravg} \fix{for the binary treatment variable~\cite{austin2017estimating, austin2022bootstrap}}.\footnote{\url{https://ngreifer.github.io/WeightIt/articles/estimating-effects.html\#binary-outcomes}} We observe that estimated odds ratio for the treatment variable (with Copilot for PRs) is 1.57, with a 95\% confidence interval ranging from 1.35 to 1.84. The \texttt{p-value} associated with this estimate was less than 0.001, indicating statistical significance at conventional alpha levels. This result implies that Pull Requests generated with the aid of Copilot for PRs are approximately 1.57 times more likely to be merged than those created without it. Given the narrow confidence interval and the statistical significance of the estimate, we can confidently assert that the treatment variable (with Copilot for PRs) exerts a meaningful impact on the likelihood of a Pull Request being merged.

\begin{tcolorbox}
\textbf{RQ2 Summary:} 
Using Copilot for PRs has a positive impact on reducing review time, reducing it by 19.3 hours. It also has a positive impact on the likelihood of a Pull Request being merged, with a 1.57 times higher chance of being merged compared to PRs whose descriptions were not generated by Copilot for PRs.
\end{tcolorbox}

\subsection{RQ3: How do developers adapt the content suggested by Copilot for PRs?}

\begin{table}[t]
\caption{Definition and frequency of complementary information categories.}
\centering
\label{tab:code1}
\resizebox{\columnwidth}{!}{
\begin{tabular}{p{40mm}p{100mm}r@{}r}
\toprule
\textbf{Category} & \textbf{Definition} &  \multicolumn{2}{c}{\textbf{Frequency}} \\ 
\midrule

Static Template Information & Pre-existing content that resides in the PR template and does not require modification, except for checklists. & 137 & (22.8\%) \\
Associated Link & Reference link or identifier corresponding to a related issue, Pull Request, or documentation. & 136 & (22.7\%) \\
Pull Request Intent & A brief description outlining the objectives or intention of the PR. & 77 & (12.8\%) \\
Testing Information & Procedures and results of testing. & 55 & (9.2\%) \\
Custom Changelog & A developer-defined changelog. & 46 & (7.7\%) \\
Visual Representation & Graphical elements (e.g., image) that provide visual context (e.g., diagram, testing screenshot, change screenshot) for the changes. & 44 & (7.3\%) \\
Future Tasks & A list of tasks or objectives to be completed in the future. & 13 & (2.2\%) \\
Code Snippet & Code excerpts that illustrate specific changes implemented in the PR or test scripts. & 8 & (1.3\%) \\
Reproduction Steps & Detailed step-by-step guide on replication of the observed behavior or issue. & 5 & (0.8\%) \\

Authorship & Certification to prove the contributor has signed off on the contribution. & 5 & (0.8\%) \\

Execution Log & Compiled log files representing the outcome of code execution or test runs. & 4 & (0.7\%) \\

Performance Impact & Impact of the PR on the software's performance metrics. & 4 & (0.7\%) \\
No Information & No supplementary information is provided. & 66 & (11.0\%) \\
\bottomrule
\end{tabular}}
\end{table}

From 13 categories of complementary information in Table~\ref{tab:code1} and seven main categories of editorial actions in Table~\ref{tab:code2}, we observe that the common category for complementary information is static template information, comprising 22.8\% of the total. Closely following is the associated link at 22.7\%. Meanwhile, both the execution log and performance impact are notably less common, each making up 0.7\%. In terms of editorial actions, deletions emerge as the most common at 22.9\%, with refinement and exclusion also being significant at no less than 17.4\%. Augmentation, on the other hand, is relatively infrequent at 6\%. In the following, we show examples of the most common and worth-noting categories.

\RqThreeDotOne
The most prevalent form of additional information (accounting for 22.8\%) is static template information.
While these PR templates are not our central concern, we advocate for a more in-depth exploration of this facet, as detailed by \citet{li2022follow}. We also found some GitHub repositories used the Copilot for PRs marker tags in their Pull Request template.\footnote{\url{https://github.com/vlang/v/blob/master/.github/PULL_REQUEST_TEMPLATE\#L47}} The second most frequent category of additional content is associated links, which serve to reference related software artifacts. We provide \fix{the example\footnote{\url{https://github.com/emqx/emqx/pull/11276}}} below for both static template information and associated links.

\begin{markdowncode}
Fixes #10647 
<details>
Copilot Summary
</details>

## Checklist for CI (.github/workflows) changes
- [ ] If changed package build workflow, pass [this action](https://github.com/emqx/emqx/actions/workflows/build_packages.yaml) (manual trigger)
- [ ] Change log has been added to `changes/` dir for user-facing artifacts update

\end{markdowncode}
Developers sometimes augment the PR descriptions with contextual information that Copilot for PRs fails to capture, such as the intent behind the PR, testing, and images. Notably, some developers even maintain their own changelogs within PRs, accounting for 7.7\% of the PRs. In the example\fix{\footnote{\url{https://github.com/owid/owid-grapher/pull/2213}}} below, the developer listed the changes to complement the content suggested by Copilot for PRs.

\begin{markdowncode}
- Lifts gdoc state out of the preview page component and puts it into the `GdocsStore`
- Restyles index page
- Adds search

<details>
Copilot Summary
</details>
\end{markdowncode}

\begin{table}[t]
\caption{Definition and frequency of editorial action categories.}
\centering
\label{tab:code2}
\resizebox{\columnwidth}{!}{
\begin{tabular}{p{70mm}p{60mm}p{6mm}@{}r}
\toprule
\textbf{Category} & \textbf{Definition} &  \multicolumn{2}{c}{\textbf{Frequency}} \\ 
\midrule
\rowcolor{grayrow}
\textbf{Deletion} & \multirow[t]{10}{\linewidth}{Eliminate the Copilot-generated content by majorly (i.e., at least two sentences) or partially (i.e., up to one sentence). Additionally, the developer sometimes eliminates any referenced URLs in the Walkthrough or drops the header generated by Copilot for PRs.} & \textbf{95} & \textbf{22.9\%} \\
Partial Summary Deletion &  & 33 & 8.0\% \\
Eliminate Walkthrough Bullet Point &  & 27 & 6.5\% \\
Major Summary Deletion &  & 14 & 3.4\% \\
Remove Copilot Header &  & 9 & 2.2\% \\
Partial Walkthrough Deletion &  & 5 & 1.2\% \\
Omit Diff Links in Walkthrough &  & 4 & 1.0\% \\
Major Walkthrough Deletion &  & 1 & 0.2\% \\
Major Poem Deletion &  & 1 & 0.2\% \\
Partial Poem Deletion &  & 1 & 0.2\% \\
\midrule
\rowcolor{grayrow}
\textbf{Refinement} & \multirow[t]{7}{\linewidth}{Refine the Copilot-generated content by majorly (i.e., at least two sentences) or minorly (i.e., up to one sentence).} & \textbf{86} & \textbf{20.8\%} \\   
Minor Summary Refinement &  & 48 & 11.6\% \\
Minor Walkthrough Refinement &  & 17 & 4.1\% \\        
Major Summary Refinement &  & 16 & 3.9\% \\
Minor Poem Refinement &  & 3 & 0.7\% \\        
Major Walkthrough Refinement &  & 2 & 0.5\% \\
\midrule
\rowcolor{grayrow}
\textbf{Exclusion} & \multirow[t]{5}{\linewidth}{Exclude the entire Copilot-generated content, the extraneous characters added before and after marker tags by developers, or the duplicate content generated by Copilot for PRs.} & \textbf{72} & \textbf{17.4\%}  \\
Exclude Poem &  & 34 & 8.2\% \\
Exclude Walkthrough &  & 22 & 5.3\% \\
Exclude Summary &  & 10 & 2.4\% \\
Remove Developer-Added Extraneous Characters &  & 5 & 1.2\% \\
Remove Duplicate Copilot Content &  & 1 & 0.2\% \\
\midrule
\rowcolor{grayrow}
\textbf{Replacement} & \multirow[t]{6}{\linewidth}{Substitute the Copilot-generated content (including the header or emoji) to a developer-defined content.} & \textbf{61} & \textbf{14.7\%} \\   
Replace Summary &  & 46 & 11.1\% \\
Replace Copilot Header &  & 11 & 2.7\% \\
Replace Walkthrough &  & 3 & 0.7\% \\        
Replace Link in Walkthrough &  & 1 & 0.2\% \\
\midrule
\rowcolor{grayrow}
\textbf{Exchangement} & \multirow[t]{10}{\linewidth}{Switch/rearrange the Copilot-generated content by the need.}  & \textbf{51} & \textbf{12.3\%} \\  
Rearrange Copilot-Generated Content &  & 18 & 4.3\% \\
Switch Summary to Comprehensive Content &  & 13 & 3.1\% \\
Switch Walkthrough to Comprehensive Content &  & 10 & 2.4\% \\
Switch Comprehensive Content to Summary &  & 3 & 0.7\% \\
Switch Walkthrough to Summary &  & 2 & 0.5\% \\
Switch Summary to Poem &  & 2 & 0.5\% \\
Switch Summary to Walkthrough &  & 1 & 0.2\% \\
Switch Poem to Walkthrough &  & 1 & 0.2\% \\
Switch Poem to Comprehensive Content &  & 1 & 0.2\% \\
\midrule
\rowcolor{grayrow}
\textbf{Augmentation} & \multirow[t]{4}{\linewidth}{Incorporate additional changes that Copilot for PRs failed to include in the Copilot-generated content, and context information, e.g., the PR impact, PR intent, and explanation for Copilot for PRs emoji.} & \textbf{25} & \textbf{6.0\%} \\
Augment Summary &  & 19 & 4.6\% \\
Augment Walkthrough & & 4 & 1.0\% \\
Add explanations to Copilot emoji & & 2 & 0.5\% \\
&&&\\
        
\midrule
\rowcolor{grayrow}
        \textbf{False Positive} & Incorrectly categorized as edited. & \textbf{24} & \textbf{5.8\%} \\
\bottomrule
\end{tabular}}
\end{table}

\RqThreeDotTwo
 Deletion emerges as the most frequent editorial action (22.9\%), which indicates developers partially choose the Copilot-generated content based on their needs. For instance,\fix{\footnote{\url{https://github.com/autowarefoundation/autoware.universe/pull/3369}}} one developer removed superfluous PR improvement statements from a Copilot-generated summary (text with a strikethrough indicating content that has been deleted).

\begin{tcolorbox}[colback=gray!5!white, arc=0pt, outer arc=0pt, colframe=gray!30!black, boxsep=5pt, boxrule=0.5pt]\small
\twemoji[width=1em]{robot} Generated by Copilot at 7a46a2f
 
Reduced the log level of some messages in util.cpp to avoid cluttering the output with non-critical information. \sout{This improves the readability and performance of the lane change planner.}
\end{tcolorbox}
It is worth noting that some PR authors opt to eliminate all links to code diffs, presumably to maintain the aesthetic integrity of the PR. In the example\fix{\footnote{\url{https://github.com/trezor/trezor-suite/pull/7962}}} below, the developer removed both the link references and a change item represented as a bullet point in the Walkthrough section.

\begin{tcolorbox}[colback=gray!5!white, arc=0pt, outer arc=0pt, colframe=gray!30!black, boxsep=5pt, boxrule=0.5pt]\small
\twemoji[width=1em]{robot} Generated by Copilot at 564357f
 \begin{itemize}
 \item Add a production debug feature that allows the user to enable the dev utils button by tapping seven times on the commit hash text in the settings screens \sout{(link, link, link, link, link, link)}
 \item \sout{Add a CopyLogsButton component that allows the user to copy the application logs to the clipboard (link, link)}
\end{itemize}
\end{tcolorbox}

 Additionally, developers often fine-tune the phrasing or terminology used by Copilot for PRs, such as correcting variable names. Below, we provide an example\fix{\footnote{\url{https://github.com/dotCMS/core/pull/25745}}} where the variable name `BASIC\_METADATA\_OVERRIDE\_KEYS' was corrected to `BASIC\_METADATA\_KEYS' (text with a strikethrough indicating content that has been deleted, while bold text indicating content that has been added).

\begin{tcolorbox}[colback=gray!5!white, arc=0pt, outer arc=0pt, colframe=gray!30!black, boxsep=5pt, boxrule=0.5pt]\small
\twemoji[width=1em]{robot} Generated by Copilot at 3454d39
 
This pull request adds a new feature that allows customizing the file asset metadata fields that are exposed by the API. It introduces a new configuration property \sout{`BASIC\_METADATA\_OVERRIDE\_KEYS'}\textbf{`BASIC\_METADATA\_KEYS'} and modifies the `FileMetadataAPIImpl' class to use it.
\end{tcolorbox}

\clearpage
Another example\fix{\footnote{\url{https://github.com/wandb/wandb/pull/5940}}} of a refinement is shown in the following. The developer rephrased a term to describe changes in a PR (i.e., `unused methods' to `duplicate method definitions').

\begin{tcolorbox}[colback=gray!5!white, arc=0pt, outer arc=0pt, colframe=gray!30!black, boxsep=5pt, boxrule=0.5pt]\small
\twemoji[width=1em]{robot} Generated by Copilot at 8a75e3a 

This Pull Request removes unnecessary quotes around type annotations in various files and classes, following the PEP 484 style guide for type hints. This improves the readability and consistency of the code and avoids the need for forward references. It also removes some \sout{unused methods}\textbf{duplicate method definitions} from the WandbCallback class and simplifies the string formatting in the results_data_frame function.
\end{tcolorbox}

Moreover, developers prune extraneous material from the \texttt{copilot:all} tag, with 11\% specifically removing \texttt{copilot:poem}. They also replace Copilot-generated content with their own (14.7\%) or exchange Copilot-generated content for another type of Copilot-generated content (12.3\%).
As shown in the example\fix{\footnote{\url{https://github.com/unoplatform/uno/pull/12596}}} below, the developer commented on the summary by reformatting it as strikethrough to abort this content suggested by Copilot for PRs.

\begin{tcolorbox}[colback=gray!5!white, arc=0pt, outer arc=0pt, colframe=gray!30!black, boxsep=5pt, boxrule=0.5pt]\small
\twemoji[width=1em]{robot} Generated by Copilot at b807adc
 
\sout{This Pull Request adds support for the Skia backend to the SamplesApp project, refactors the rendering and extension logic of the Uno.UI.Composition and Uno.UI.Runtime.Skia.Gtk projects, and aligns the app manifest of the SamplesApp.UWP project with the Uno Platform branding. It also renames, deletes, or updates some files and namespaces to improve the code organization and consistency.}

Nope, you didn't get it this time
\end{tcolorbox}

In 6\% of instances, developers add additional changes or context information to augment the content suggested by Copilot for PRs.
In the following example,\fix{\footnote{\url{https://github.com/owid/owid-grapher/pull/2109}}} the developer added hidden emoji explanations (which were commented out in the PR description) and replaced the summary with a developer-defined summary.

\begin{tcolorbox}[colback=gray!5!white, arc=0pt, outer arc=0pt, colframe=gray!30!black, boxsep=5pt, boxrule=0.5pt]\small
\twemoji[width=1em]{bug}\twemoji[width=1em]{bar chart}\twemoji[width=1em]{rainbow}  \textbf{<-- copilot thought of this (and also gave an explanation, see below)}\\

\sout{Fix highlighting bugs in StackedAreaChart and StackedBarChart when using external legends. Use rawSeries instead of series to match the legend names with the chart data.}
\\
\textbf{Fixes a problem in StackedArea and StackedBar charts where hovering over an entity in the legend didn't grey out all other entities. Here, Italy is hovered over:}

\textbf{Emoji legend (generated by copilot):}

\twemoji[width=1em]{bug} \textbf{- This emoji represents a bug fix, which is the main purpose of these changes. The bug was causing the wrong series to be highlighted on the stacked area and bar charts when using an external legend.}

\twemoji[width=1em]{bar chart} \textbf{- This emoji represents a chart or graph, which is the type of component that these changes affect. The stacked area and bar charts are both chart components that use an external legend to display the series names and colors.}

\twemoji[width=1em]{rainbow} \textbf{- This emoji represents a rainbow or color, which is a relevant aspect of these changes. The highlighting feature of the external legend and the chart components depends on the color of the series, and these changes ensure that the correct color is used for the correct series.}
\end{tcolorbox}

Lastly, we identified 24 false positives in our dataset, occurring when Copilot for PRs and developers edited the PR description simultaneously, leading to content reappearances,\footnote{\url{https://github.com/dotCMS/core/pull/25770}} or when the developer deleted the edit history,\footnote{\url{https://github.com/pancakeswap/pancake-frontend/pull/7173}} or when Copilot for PRs restored the edited PR template to its original state.\footnote{\url{https://github.com/lensterxyz/lenster/pull/2413}}

\begin{tcolorbox}
\textbf{RQ3 Summary:} 
We identified 13 categories of supplementary information and observed seven edit actions to content suggested by Copilot for PRs. For strategies of adapting AI generated content, developers tend to include templates and associated links to complement it. Moreover, they often partially removed content suggested by Copilot for PRs.  
\end{tcolorbox}

\section{Discussion}
We now discuss the recommendations from our results, as well as the threats to the validity of our study.

\subsection{Recommendations}
Based on our findings, we make the following recommendations for practitioners, researchers, and GitHub Copilot for PRs. First,
we recommend that practitioners:

\begin{itemize}
    \item \textit{Advocate for Copilot for PRs adoption in PRs:} According to our results (\textbf{RQ2}), PRs enriched by Copilot for PRs generally necessitate less review time and exhibit an increased likelihood of being merged. We champion the utilization of AI-powered descriptions to amplify the clarity of the changes outlined in PRs.
    
    \item \textit{Incorporate Copilot for PRs tags into PR templates:} Our data reveals occasional developer inclusion of Copilot for PRs tags in PR templates. Additionally, templates often act as a supplemental layer to Copilot-generated content (\textbf{RQ3.1}). Thus, we advise practitioners to harmonize their unique PR requirements with these AI-generated tags.
    
    \item \textit{Exercise discretion with the \texttt{copilot:all} tag:} Our \textbf{RQ3.2} results indicate that developers frequently exclude the `poem' section when employing the \texttt{copilot:all} tag. To minimize confusion for reviewers, it is advisable to be selective when including various types of Copilot-generated content.
\end{itemize}

For researchers, we suggest:
\begin{itemize}
    \item \textit{Examine content evolution across commits:} As noted in Section~\ref{sec:edit}, developers often redeploy Copilot for PRs tags to assess generated content against new commits. A nuanced qualitative analysis of such behavior poses a prospective research direction.
    
    \item \textit{Establish Copilot for PRs tag integration guidelines:} Although we advocate for the incorporation of Copilot for PRs tags into PR templates, devising guidelines based on the specific categories behind PR templates~\cite{li2022follow} could prove beneficial.
\end{itemize}

Future research directions that could augment GitHub Copilot for PRs include:
\begin{itemize}
    \item \textit{Refinement of template integration:} Given that templates frequently complement Copilot-generated content (\textbf{RQ3}), Copilot for PRs could benefit from offering more customized or comprehensive content tailored to specific repositories.
    
    \item \textit{Offer developer-specific suggestions:} Our findings in \textbf{RQ3.2} indicate a propensity for developers to tailor suggestions from Copilot for PRs. Allowing for such customization could further enhance the utility of Copilot for PRs.
    
    \item \textit{Learn from Developer Modifications:} As observed in \textbf{RQ3.2}, developers frequently modify, remove, or augment specific elements within suggestions from Copilot for PRs. Capturing these alterations could provide valuable learning opportunities for enhancing the platform's generated content.
\end{itemize}

\subsection{Threats to validity}
Below, we outline potential threats to the validity of our study:

\textbf{Construct Validity:} This study focuses on how developers adopt Copilot for PRs, thereby excluding PRs submitted by bots. We identify bots based on a ``bot'' suffix and employ techniques by \citet{golzadeh2022accuracy}. Although these methods are highly precise, they might not capture every true negative, potentially influencing the construct validity. 

\textbf{Internal Validity:} Our study hinges on manually coded data, which is susceptible to miscoding due to the subjective interpretations of the coding schema. To counteract this, we adhere to two best practices for open coding: 1) we execute two rounds of independent coding and compute the Kappa statistic to assure a `Substantial' level of agreement, and 2) if the coding schema undergoes modifications, we revisit and adjust the existing coding accordingly. Additionally, we construct our quasi-experiments considering 17 confounding variables (as detailed in Table~\ref{tab:variables}). These variables have been shown to have an influence on review time, outcome, quality, and participation. However, there may be other confounding variables not accounted for, necessitating further analysis. \fix{Furthermore, we assessed the balance achieved by propensity score weighting, focusing on quality assessment through this method rather than traditional fit statistics like R-squared. This approach aligns with the perspective that traditional goodness-of-fit tests are of limited relevance in this context. The primary goal of weighting is to balance covariate distributions within the sample, rather than to infer assignment probabilities in the overall population~\cite{li2018balancing}.}

\textbf{External Validity:} GitHub Copilot for PRs was introduced in March 2023, making limited users early adopters. Given the limited number of developers who have access at this stage, there is an inherent threat to the external validity of our empirical findings. As such, it is crucial to clarify that our results are not universally applicable to the broader open-source developer community, but are more pertinent to these early adopters. \fix{Developers who are less eager to adopt new technologies might use Copilot for PRs less or differently compared to the early adopters studied in this work.}

\section{Conclusion}


In this work, we examined how developers are using generative AI for writing and reviewing Pull Requests (PRs) and its effects on the code review process. Our study included over 18,000 PRs assisted by Copilot for PRs and more than 54,000 that were not. We found a growing trend of Copilot for PRs use in code reviews. Some repositories have fully embraced it, while others are still testing the waters.

Our data shows that Copilot for PRs can reduce the time needed for code reviews and increase the chances of PRs getting merged. Developers are also augmenting the AI-generated content, demonstrating a unique interplay where human expertise edits and refining the machine-generated suggestions to ensure contextual relevance and technical accuracy.

Looking ahead, our exploration into generative AI for PR descriptions is just the beginning of exploring the potential for human-AI collaboration in the context of code reviews and other software development tasks. We envision future work on empirically exploring the adoption of AI and the adaptive strategies employed by developers to tailor AI outputs in the areas of code reviews, code creation, and documentation, among others.

\section*{Data Availability}
Our replication package~\cite{replicate} includes lists of studied PRs from GitHub, both with and without the use of Copilot for PRs. It also provides the features of PRs that were either generated or not generated by Copilot for PRs (pertaining to RQ2), coding results for RQ3, and scripts. The complete replication package can be also accessed at \url{https://github.com/NAIST-SE/CopilotForPRsEarlyAdoption}.
\begin{acks}
This work was supported by JSPS Grant-in-Aid for JSPS Fellows JP23KJ1589, JSPS KAKENHI Grant Numbers JP20H05706, and JST PRESTO Grant Number JPMJPR22P6.
\end{acks}

\bibliographystyle{ACM-Reference-Format}
\bibliography{main}


\end{document}